\documentclass[sigconf,natbib=true]{acmart}
\usepackage{color}
\usepackage{bbm}
\usepackage{multirow}
\usepackage[inline]{enumitem}
\usepackage{graphicx}
\usepackage{subcaption}
\usepackage{sistyle}
\usepackage[ruled]{algorithm2e}
\usepackage{booktabs}
\usepackage{caption}
\usepackage{graphicx} 
\SIthousandsep{,}
\usepackage{makecell}
\usepackage[skip=0pt]{caption}
\usepackage{amsmath}
\usepackage{hyperref}
\usepackage{array} 
\usepackage{graphicx}
\usepackage{acronym}
\usepackage[export]{adjustbox}
\usepackage{booktabs}
\newcommand{\heading}[1]{\vspace*{0.5mm}\noindent\textbf{#1.}}
\AtBeginDocument{%
  \providecommand\BibTeX{{%
    \normalfont B\kern-0.5em{\scshape i\kern-0.25em b}\kern-0.8em\TeX}}}

\makeatletter
\g@addto@macro\normalsize{%
  \abovedisplayskip 2pt plus1pt 
  \belowdisplayskip 2pt plus1pt
  \abovedisplayshortskip  2pt plus1pt%
  \belowdisplayshortskip  1pt plus1pt
}
\makeatother

\acrodef{IR}{infor\-mation retrieval}
\acrodef{LLM}{large language model}
\acrodef{QA}{question answering}
\newcommand{\alldata}{General-Full-Data\xspace}
\newcommand{\engdata}{Eng-Text-Data\xspace}
\newcommand{\modelname}{BOOM\xspace}

\AtBeginDocument{%
  \providecommand\BibTeX{{%
    Bib\TeX}}}

\setcopyright{acmlicensed}
\copyrightyear{2018}
\acmYear{2018}
\acmDOI{XXXXXXX.XXXXXXX}

\acmConference[Conference acronym 'XX]{Make sure to enter the correct
  conference title from your rights confirmation email}{June 03--05,
  2018}{Woodstock, NY}

\acmISBN{978-1-4503-XXXX-X/18/06}
\received{20 February 2007}
\received[revised]{12 March 2009}
\received[accepted]{5 June 2009}
\author{Hengran Zhang}
\affiliation{
  \institution{State Key Laboratory of AI Safety, Institute of Computing Technology, \\
Chinese Academy of Sciences}
  \institution{University of Chinese Academy of Sciences}
  \city{Beijing}
  \country{China}
}
\email{zhanghengran22z@ict.ac.cn}

\author{Keping Bi}
\orcid{}
\affiliation{
	 \institution{State Key Laboratory of AI Safety, Institute of Computing Technology, \\
Chinese Academy of Sciences}
  \institution{University of Chinese Academy of Sciences}
	\city{Beijing}
	\country{China}
}
\email{bikeping@ict.ac.cn}

\author{Jiafeng Guo}
\affiliation{
	 \institution{State Key Laboratory of AI Safety, Institute of Computing Technology, \\
Chinese Academy of Sciences}
  \institution{University of Chinese Academy of Sciences}
	\city{Beijing}
	\country{China}
}
\email{guojiafeng@ict.ac.cn}

\author{Jiaming Zhang, Wenbo Yang}
\affiliation{
	\institution{Querit Private Limited}
 \city{Singapore}
 \country{}
}
\email{jm.zhang@querit.ai}
\email{bob@querit.ai}

\author{Daiting Shi}
\affiliation{
	\institution{Querit Private Limited}
 \city{Singapore}
 \country{}
}
\email{shidaiting@querit.ai}

\author{Xueqi Cheng}
\orcid{0000-0002-5201-8195}
\affiliation{
	\institution{State Key Laboratory of AI Safety, Institute of Computing Technology, \\
Chinese Academy of Sciences}
  \institution{University of Chinese Academy of Sciences}
	\city{Beijing}
	\country{China}
}
\email{cxq@ict.ac.cn}

\begin{document}

\title{Bagging-Based Model Merging for Robust General Text Embeddings}
\begin{abstract} 
General-purpose text embedding models underpin a wide range of NLP and information retrieval applications, and are typically trained on large-scale multi-task corpora to encourage broad generalization. However, it remains unclear how different multi-task training strategies compare in practice, and how to efficiently adapt embedding models as new domains and data types continually emerge. In this work, we present a systematic study of multi-task training for text embeddings from two perspectives: data scheduling and model merging. We compare batch-level shuffling, sequential training variants, two-stage training, and multiple merging granularities, and find that simple batch-level shuffling consistently yields the strongest overall performance, suggesting that task conflicts are limited and training datasets are largely complementary. Despite its effectiveness, batch-level shuffling exhibits two practical limitations: suboptimal out-of-domain (OOD) generalization and poor suitability for incremental learning due to expensive full retraining. To address these issues, we propose Bagging-based rObust mOdel Merging (\modelname), which trains multiple embedding models on sampled subsets and merges them into a single model, improving robustness while retaining single-model inference efficiency. Moreover, \modelname naturally supports efficient incremental updates by training lightweight update models on new data with a small historical subset and merging them into the existing model. Experiments across diverse embedding benchmarks demonstrate that \modelname consistently improves both in-domain and OOD performance over full-corpus batch-level shuffling, while substantially reducing training cost in incremental learning settings.
\footnote{Our code and datasets can be found at \url{https://anonymous.4open.science/r/Bagging-Based-Model-Merging-3E60/README.md}.}

\end{abstract}


\maketitle

\section{Introduction} 
Text embeddings encode natural language into dense vectors and underpin a wide range of natural language processing (NLP) and information retrieval (IR) applications, including retrieval, reranking, classification, clustering, and semantic textual similarity (STS) \cite{izacard2021unsupervised, wang2022text, xiao2024c, muennighoff2022sgpt, neelakantan2022text, li2023towards}. They are also a core component in retrieval-augmented generation (RAG), where embedding quality directly affects the relevance of retrieved evidence and the quality of generated outputs \cite{lewis2020retrieval, gao2023retrieval}. The rapid progress of general-purpose embedding models has been accelerated by standardized evaluations, such as the Massive Text Embedding Benchmark (MTEB)\cite{muennighoff2023mteb,enevoldsen2025mmteb}, where models are expected to perform well across diverse task families and domains. A central challenge is \textbf{generalization}: representations should remain effective across both seen and unseen tasks and domains.

To promote generalization, embedding models are commonly trained on large-scale multi-task corpora spanning retrieval-style contrastive learning, supervised classification, clustering, and similarity objectives \cite{lee2024nv, lee2025gemini, li2024making, zhang2025qwen3}. They are then evaluated on both in-domain tasks and out-of-domain (OOD) benchmarks that differ in topic, domain, or task composition \cite{muennighoff2023mteb,enevoldsen2025mmteb}. Meanwhile, practical embedding systems are rarely static—new domains (e.g., legal or financial retrieval) and new data types (e.g., code retrieval) continually emerge and must be incorporated. This motivates two key research questions: \textbf{(i)} how to enable \textit{effective and efficient multi-task training} for general-purpose embeddings, and \textbf{(ii)} how to achieve \textit{effective and efficient incremental learning} without expensive full retraining.

Multi-task training is often assumed to suffer from \textit{task conflict}, where gradients from different tasks interfere and degrade performance \cite{yu2020gradient}. Consequently, prior work has studied various data scheduling strategies, including curriculum learning and sequential training, to reduce interference and improve stability \cite{DBLP:conf/icml/BengioLCW09}. In parallel, \textbf{model merging} has been proposed as an alternative paradigm: train multiple models on different datasets (or task partitions) and merge them into a single model (e.g., by weight merging or LoRA merging), aiming to combine capabilities while avoiding repeated end-to-end retraining \cite{denneng2024model}. However, it remains unclear how these approaches compare in the setting of general text embedding and whether task conflict is indeed a dominant bottleneck.

In this work, we conduct a systematic study of multi-task training strategies from two perspectives: \textit{data scheduling} and \textit{model merging}. For data scheduling, we compare batch-level shuffling, dataset-level sequential training, task-level sequential training, and two-stage training. For model merging, we evaluate dataset-level, task-level, and cluster-level merging strategies. Surprisingly, our results show that \textit{batch-level shuffling consistently achieves the strongest overall performance}, suggesting that the general text embedding tasks have limited conflicts in practice and that training datasets are largely complementary due to shared semantic matching objectives. 

\begin{figure}[!t]
    \centering
    \includegraphics[width=.8\linewidth]{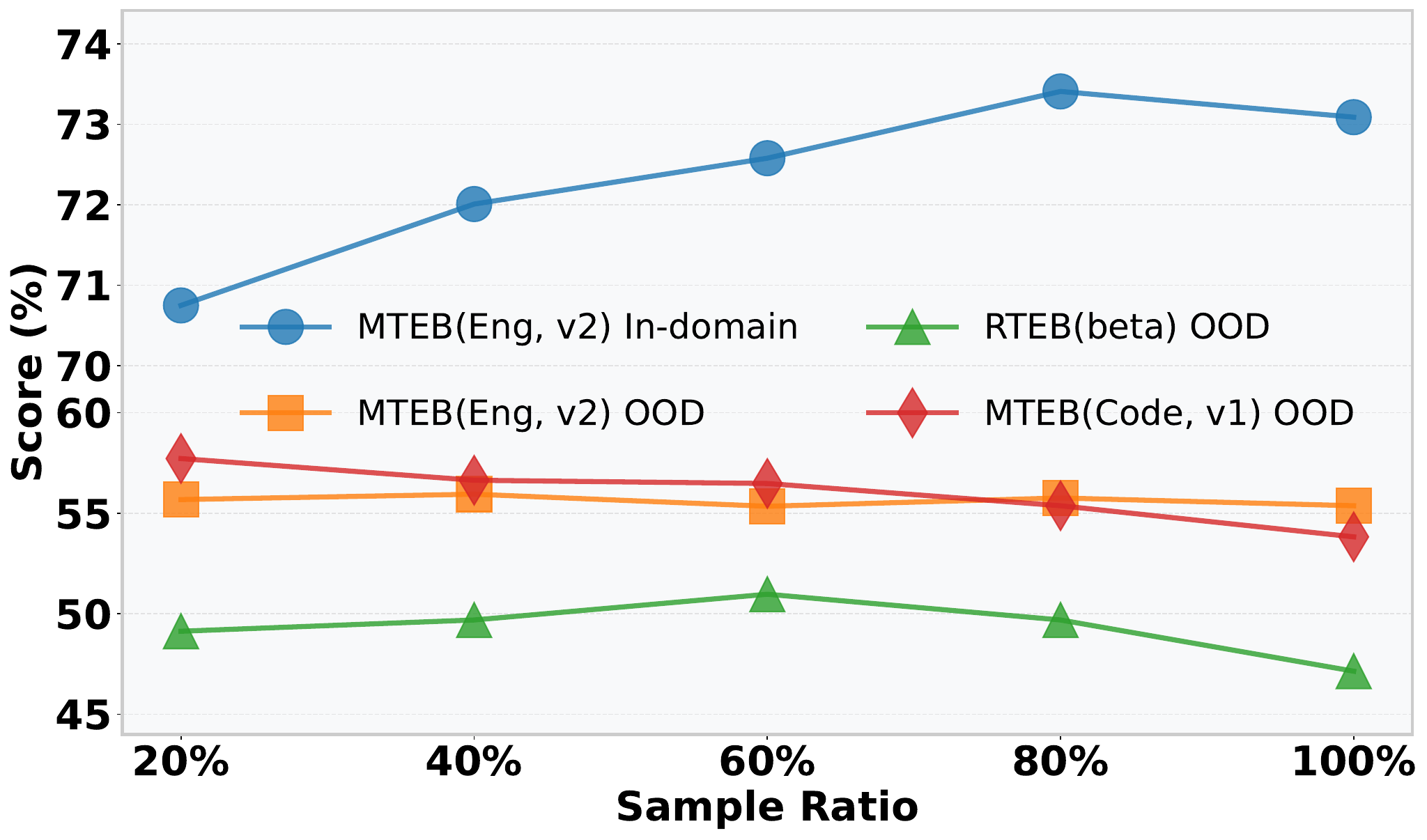}
    \caption{Average performance (\%) of general text embedding models trained with different proportions of the multi-task training set on in-domain and OOD evaluation sets.}
    \label{fig:different_ratio}
\end{figure}

However, batch-level shuffling has two practical limitations. First, it may yield suboptimal OOD generalization. As shown in Figure~\ref{fig:different_ratio}, when evaluating on OOD benchmarks, including the OOD tasks in MTEB(Eng, v2) \cite{enevoldsen2025mmteb}, domain-specific retrieval in RTEB(beta) \cite{enevoldsen2025mmteb}, and code retrieval in MTEB(Code, v1) \cite{DBLP:conf/acl/LiDL0ZDWT25}, models trained on the full dataset achieve strong in-domain performance but can underperform compared to models trained on smaller subsets (e.g., 20\%). This suggests that simply scaling training data (even of various types) does not necessarily translate into more robust generalization. Second, batch-level shuffling is poorly suited to incremental learning settings: when new data arrives, the model typically must be retrained on the entire expanded corpus, which is costly and impractical.

These observations motivate a shift from ``training on more data'' to \textit{robust generalization} under distribution shifts. A classical technique for improving robustness in machine learning is \textbf{bootstrap aggregating (bagging)}, which trains multiple models on different sampled subsets of the data and aggregates them to reduce variance \cite{DBLP:journals/ml/Breiman96b}. Yet conventional bagging requires deploying an ensemble of models, increasing inference latency and cost, undesirable for embedding services. Crucially, model merging provides a way to compress such an ensemble into a single model, retaining robustness benefits without multi-model inference.

Building on this insight, we propose \textbf{B}agging-based r\textbf{O}bust m\textbf{O}del \textbf{M}erging (\textbf{\modelname}) for general text embedding. \modelname~trains multiple embedding models on different sampled subsets using standard batch-level shuffling, and then merges them into a single model. This improves robustness and OOD generalization while preserving inference efficiency. Moreover, \modelname~naturally supports incremental learning: when new data arrives, we train a lightweight update model on the new data plus a small sampled subset of historical data, and merge it with the existing model—efficiently incorporating new knowledge while mitigating forgetting.

We evaluate both \modelname~and batch-level shuffling baselines across multiple embedding benchmarks. Experimental results show that \modelname consistently outperforms batch-level shuffling trained on the full corpus, achieving stronger performance in both in-domain and out-of-domain settings. Moreover, in incremental learning scenarios, \modelname enables efficient integration of new data via lightweight training and merging, delivering improved performance while substantially reducing training cost compared to full retraining with batch-level shuffling.

In summary, our main contributions are threefold:
1) We systematically study data scheduling and model merging paradigms for general text embedding, and find that task conflicts are limited in practice, with batch-level shuffling providing strong and consistent gains across diverse tasks.
2) We propose \modelname, which trains multiple embedding models on differently sampled subsets and merges them into a single model, improving generalization while avoiding the inference overhead of conventional ensembles.
3) We extend \modelname to incremental updates by training on new data together with a sampled historical subset and merging the resulting model with the existing one, enabling effective knowledge integration with substantially lower training cost than full retraining.

\vspace{-2mm}
\section{Related Work}

\subsection{General Text Embedding}
General text embedding represents a critical research direction in information retrieval and natural language processing, with diverse applications across web search, question answering, and retrieval-augmented generation. 
The prevailing approach employs a dual-encoder framework, in which queries and documents are encoded independently. The cosine similarity between their embeddings then serves as an estimate of semantic relevance, forming the core methodological principle behind general text embedding. 

\heading{PLM-based Embedding}
During the development of pre-trained language models (PLMs) such as BERT and T5, numerous impactful methods have been proposed to advance the use of text embeddings for general tasks. Notable examples include Contriever \cite{izacard2021unsupervised}, E5 \cite{wang2022text}, BGE \cite{xiao2024c}, SGPT \cite{muennighoff2022sgpt}, Open Text Embedding \cite{neelakantan2022text}, and GTE \cite{li2023towards}. These models can generally be categorized into two main types:
\begin{enumerate*}
\item Unsupervised and Weakly-Supervised Contrastive Learning \cite{izacard2021unsupervised, neelakantan2022text}: For example, Contriever \cite{izacard2021unsupervised} generates pseudo-positive pairs by independently cropping two distinct spans from the same document and treating them as semantically equivalent. Open Text Embedding \cite{neelakantan2022text} leverages large-scale contrastive pre-training on neighboring text pairs, which are mined from the internet (e.g., adjacent snippets).
\item Multi-Stage and Instruction-Tuned Training \cite{xiao2024c, wang2022text, li2023towards}: These models typically undergo weakly-supervised contrastive pre-training on large datasets of text pairs collected from diverse web sources (such as citation graphs and Reddit), followed by supervised fine-tuning on labeled datasets.
\end{enumerate*}

\heading{LLM-based Emebdding} Large language models (LLMs) have demonstrated strong performance across a variety of NLP tasks, and recent studies increasingly explore their potential as backbone encoders for text embedding \cite{zhang2025comparative, ma2024fine, behnamghader2024llm2vec, springer2024repetition, li2024llama2vec, zhang2025unleashing, li2025conan, lee2024nv, lee2025gemini, zhang2025qwen3}. Early work, such as Repllama \cite{ma2024fine}, first employed LLMs for embedding generation, showing substantial improvements over traditional pretrained language model (PLM)-based approaches.
However, the causal attention mechanism in decoder-only LLMs may constrain their ability to produce highly contextualized embeddings. To address this, LLM2vec\cite{behnamghader2024llm2vec} replaced causal attention with bidirectional attention and introduced a masked next-token prediction (MNTP) warm-up strategy. Similarly, Echo \cite{springer2024repetition} generated enhanced embeddings by repeating input sequences and extracting representations from the duplicated tokens. Llama2Vec \cite{li2024llama2vec} further aligned LLMs with embedding tasks through pretraining objectives tailored to representation learning, yielding strong results on the BEIR benchmark.  LLM-QL \cite{zhang2025unleashing} leverages the generative strengths of LLMs through QL maximization with Attention Block and Document Corruption, which acts as a preparation step to better warm up the model for the following contrastive training. 
Conan-Embedding-v2 \cite{li2025conan} proposes a novel soft masking mechanism combined with dynamic rank reduction.
NV-Embed \cite{lee2024nv} introduced latent attention pooling and Two-stage training for state-of-the-art representation quality. 
Gemini embedding \cite{lee2025gemini} employed two-stage training, i.e., pre-finetuning on larger scale weak supervied data and fine-tuning on high-quality supervised data.  
Qwen3-Embedding \cite{zhang2025qwen3} integrated large-scale LLM-generated synthetic data, multi-stage fine-tuning, and model merging to deliver better embeddings. 
While its merging strategy follows \citet{DBLP:journals/corr/abs-2410-15035}, which proposed the model merging to address task conflicts during general text embedding model training. 
Unlike \citet{DBLP:journals/corr/abs-2410-15035}, we conduct a more comprehensive and rigorous investigation of various training strategies for general text embedding. 


\subsection{Ensemble Learning}
Ensemble learning \cite{DBLP:journals/fcsc/DongYCSM20, DBLP:journals/widm/SagiR18} is a powerful machine learning paradigm where multiple models (often referred to as ``base learners'') are combined to solve a problem and produce a better performance than individual models. 
Bootstrap Aggregating (Bagging) is one of the most popular ensemble learning techniques that combines multiple models to improve predictive accuracy, especially for high-variance, low-bias models like decision trees. 
Introduced by \cite{DBLP:journals/ml/Breiman96b}, the core idea of bagging is to reduce variance and avoid overfitting by aggregating the predictions of several models trained on different subsets of the data. 
Each model is trained on a bootstrapped sample, which is a randomly chosen subset of the training data with replacement. 
However, the high computational and inference cost of bagging remains a significant drawback, especially for tasks requiring real-time predictions using large models, e.g., LLMs. 
To address this, model merging offers promising solutions, allowing for faster inference by consolidating multiple models into a single one. 
These methods can make bagging more feasible for applications requiring low-latency predictions, such as online text embedding, and can help overcome the challenges posed by multi-model inference in practice.

\vspace{-2mm}
\section{Preliminary}
Model merging is a powerful technique that combines the strengths of multiple models without incurring the computational overhead of ensembling or the need for additional training. MergeKit \cite{goddard-etal-2024-arcees} is a comprehensive open-source library that streamlines the implementation of various model-merging strategies. Given a base model $W_0$ and $N$ fine-tuned models $\{W_1, W_2, \ldots, W_N\}$, these $N$ models can be merged into a single model, $\mathbf{W}_{\text{merged}}$.  
The following section outlines the key merging approaches available in MergeKit.

\subsection{Spherical Interpolation Methods}
Spherical interpolation methods are designed to merge or interpolate model weight vectors while respecting their geometric structure on the hypersphere.  
\begin{itemize}[leftmargin=*,itemsep=0pt,topsep=0pt,parsep=0pt]
\item \textbf{Spherical Linear Interpolation (SLERP)}. SLERP \cite{shoemake1985animating} interpolates between two model weight vectors $\mathbf{W}_A$ and $\mathbf{W}_B$ on the hypersphere using
\begin{equation}
\mathbf{W}_{\text{slerp}} = \frac{\sin((1-\alpha)\theta)}{\sin\theta}\mathbf{W}_A + \frac{\sin(\alpha\theta)}{\sin\theta}\mathbf{W}_B,
\end{equation}
where $\theta = \arccos\left(\frac{\mathbf{W}_A \cdot \mathbf{W}_B}{\|\mathbf{W}_A\|\|\mathbf{W}_B\|}\right)$ and $\alpha \in [0,1]$.
SLERP is inherently a pairwise interpolation method, which means it can directly merge only two models at a time. To merge more than two models ($K>2$), a common strategy is to recursively apply SLERP in a sequential fashion: repeatedly merge pairs of models until a final merged model $\mathbf{W}_{\text{merged}}$ is obtained. While simple, this sequential approach may be sensitive to the order in which models are merged.  

\item \textbf{Multi-SLERP} \cite{goddard2024arcee}. Multi-SLERP generalizes SLERP to $N$ models $\{\mathbf{W}_i\}_{i=1}^N$ with barycentric weights $\{\alpha_i\}_{i=1}^N$ ($\sum_{i=1}^N \alpha_i=1$) as follows:
\begin{equation}
\mathbf{W}_{\text{merged}} = \left(\sum_{i=1}^N \alpha_i \|\mathbf{W}_i\|\right) \cdot \exp_{\mathbf{M}}\left(\sum_{i=1}^N \alpha_i \log_{\mathbf{M}}\left(\frac{\mathbf{W}_i}{\|\mathbf{W}_i\|}\right)\right),
\end{equation}
where $\mathbf{M}$ is the normalized weighted mean direction:
\begin{equation}
\mathbf{M} = \frac{\sum_{i=1}^N \alpha_i \frac{\mathbf{W}_i}{\|\mathbf{W}_i\|}}{\left\|\sum_{i=1}^N \alpha_i \frac{\mathbf{W}_i}{\|\mathbf{W}_i\|}\right\|}.
\end{equation}
Here, $\log_{\mathbf{M}}(\cdot)$ and $\exp_{\mathbf{M}}(\cdot)$ denote the logarithmic and exponential maps at $\mathbf{M}$ on the sphere.

\item \textbf{Karcher Mean} \cite{goddard2024arcee}.  
The merged model $\mathbf{W}_{\text{merged}}$ is defined as the point on the hypersphere that minimizes the sum of squared geodesic distances to the input vectors. Formally,
\begin{equation}
\mathbf{W}_{\text{merged}} = \arg\min_{\mathbf{W} \in \mathcal{S}} \sum_{i=1}^N d^2(\mathbf{W}, \mathbf{W}_i),
\end{equation}
where $\mathcal{S}$ denotes the unit hypersphere and $d(\cdot, \cdot)$ represents the geodesic (angular) distance between two points on the sphere. In practice, computation of the Karcher mean is typically performed via an iterative optimization procedure. This approach produces a consensus model that faithfully captures the central tendency of the input models in the underlying spherical geometry.  
\end{itemize}

\subsection{Task Vector Based Methods}
The following methods build on the notion of ``task vectors'', defined as the difference between a fine-tuned model and its base model.
\begin{itemize}[leftmargin=*,itemsep=0pt,topsep=0pt,parsep=0pt]
\item \textbf{Task Arithmetic}. Task Arithmetic \cite{ilharco2022editing} defines task vectors $\mathbf{\tau}_i = \mathbf{W}_i - \mathbf{W}_{\text{base}}$ and combines them with weights $\alpha_i$:
\begin{equation}
\mathbf{W}_{\text{merged}} = \mathbf{W}_{\text{base}} + \sum_{i=1}^N \alpha_i \mathbf{\tau}_i.
\end{equation}
\item \textbf{TIES Merging}. TIES \cite{yadav2023ties} also uses task vectors, but attempts to mitigate conflicts between merges in weight space. TIES proceeds in three stages: (1)Trim: For each layer, retain only the top-$k\%$ largest-magnitude entries in $\mathbf{\tau}_i$, set the rest to zero: $\mathbf{\tau}_i' = \text{TopK}(\mathbf{\tau}_i, k)$; (2) Sign Selection: Compute the sign consensus for each parameter across all $\mathbf{\tau}_i'$. Mask out parameters that disagree with the consensus: $\mathbf{m}_{\text{consensus}} = \text{sign}\left(\sum_{i=1}^N \alpha_i \mathbf{\tau}_i'\right)$; (3) Merge: 
\begin{equation}
    \mathbf{W}_{\text{merged}} = \mathbf{W}_{\text{base}} + \frac{\sum_{i=1}^N \alpha_i (\mathbf{\tau}_i' \odot \mathbf{m})}{\sum_{i=1}^N \alpha_i \mathbf{m}}
\end{equation}
\item \textbf{SCE}. Unlike TIES, which first performs layer-wise magnitude pruning (TopK) before consensus filtering, SCE \cite{wan2025fusechat} operates on the global sign consistency directly. For each parameter, retain only entries where all task vectors share the same sign, and mask out conflicting dimensions:
\begin{equation}
\mathbf{m}_\text{sign} = \mathbb{I}\left(\text{AllSameSign}\left(\{\boldsymbol{\tau}_i\}_{i=1}^N\right)\right), \boldsymbol{\tau}_i' = \boldsymbol{\tau}_i \odot \mathbf{m}_\text{sign},
\end{equation}
and then merge:
The surviving updates are then aggregated via a weighted sum and normalization:
\begin{equation}
    \mathbf{W}_\text{merged} = \mathbf{W}_\text{base} +
\frac{\sum_{i=1}^{N} \alpha_i \boldsymbol{\tau}i’}{\sum{i=1}^N \alpha_i \mathbf{m}_\text{sign}}
\end{equation}


\end{itemize}

\subsection{Specialized Methods}
\textbf{Model Stock}. Model Stock \cite{jang2024model} moves the merged weights toward the geometric center of a set of fine-tuned checkpoints. Specifically, model stock computes optimal interpolation between a base model $\mathbf{W}_0$ and the average of fine-tuned models $\overline{\mathbf{W}}$. 
Using the average cosine similarity $\overline{\cos\theta}$ between task vectors: 
\begin{equation}
t = \frac{N\overline{\cos\theta}}{1 + (N-1)\overline{\cos\theta}}, \quad
\mathbf{W}_{\text{merged}} = t\overline{\mathbf{W}} + (1-t)\mathbf{W}_0. 
\end{equation}


\section{Experimental Setup}

\subsection{Training Data}
Our training framework utilizes two primary datasets: the specialized \engdata and a broader, more comprehensive collection termed \alldata, which encompasses multilingual text retrieval, code retrieval, and additional diverse training sources.

\heading{\engdata}
We adopt the extensively curated dataset introduced by \citet{li2024making}, which integrates multiple publicly available benchmarks across several key tasks:
\begin{itemize}[leftmargin=*,itemsep=0pt,topsep=0pt,parsep=0pt]
\item \textbf{Retrieval:} ELI5 \cite{fan2019eli5}, HotpotQA \cite{yang2018hotpotqa}, FEVER \cite{thorne2018fever}, MSMARCO passage and document ranking \cite{bajaj2016ms}, NQ \cite{kwiatkowski2019natural}, NLI, SQuAD \cite{rajpurkar2016squad}, TriviaQA \cite{joshi2017triviaqa}, and FiQA \cite{maia201818}.
\item \textbf{Reranking:} StackOverFlowDupQuestions \cite{liu2018linkso}.
\item \textbf{Classification:} AmazonReviews-Classification \cite{mcauley2013hidden},  Banking77-Classification \cite{casanueva2020efficient}, Emotion-Classification \cite{saravia2018carer},  MTOPIntent-\\Classification \cite{li2021mtop}, IMDB-Classification \cite{maas2011learning}, ToxicConversations-Classification \cite{adams2019jigsaw}, TweetSentimentExtraction-Classification \cite{wei2020tweet}, \\ AmazonCounterfactual-Classification \cite{o2021wish}.
\item \textbf{Clustering:} {Arxiv/Biorxiv/Medrxiv/Reddit/StackExchange}-\\Clustering-{S2S/P2P}, TwentyNewsgroups-Clustering \cite{lang1995newsweeder}.
\item \textbf{Semantic Text Similarity (STS):} STS12 \cite{agirre2012semeval}, STS22 \cite{chen2022semeval}, STS-Benchmark \cite{cer2017semeval}.
\end{itemize}
The original training data provided by BGE-en-ICL \cite{li2024making} includes three datasets—Quora Duplicate Questions \cite{thakur2021beir}, SCIDOCS-RR \cite{specter2020cohan}, and ArguAna~\cite{wachsmuth2018retrieval}—for which MTEB provides only test or development splits, but no training data. To avoid potential data contamination, we exclude these datasets from \engdata.  
In total, this comprehensive training corpus comprises 31 distinct datasets, encompassing approximately 2M data points.

\heading{\alldata}
In addition to the English in-context learning data above, we expand the training corpus with multilingual retrieval datasets to enhance cross-lingual generalization. These include DuReader \cite{he2018dureader}, MIRACL \cite{zhang2023miracl}, Mr. TyDi \cite{zhang2021mr}, and T2-Ranking \cite{xie2023t2ranking}. Furthermore, to support code retrieval capabilities, we incorporate code-specific training examples sourced from \citet{suresh2024cornstack}. Specifically, we sample approximately 10,000 training queries for each of the following programming languages: JavaScript, Java, Python, PHP, and Ruby. This combined dataset ensures coverage across diverse languages, formats, and task types, contributing to a robust and generalizable model training process. 
In total, General-Full-Data comprises approximately 2.8M data points. 

\vspace{-2mm}
\subsection{Evaluation Setting}
To rigorously assess the model's capabilities, we evaluate its performance across a suite of established benchmarks from the Massive Text Embedding Benchmark (MTEB) framework \cite{muennighoff2023mteb}.

\heading{MTEB (English, v2)} This updated English benchmark offers a more realistic assessment of model generalization compared to its predecessor (v1). Its scope includes 41 datasets across 7 task types (i.e., retrieval, classification, clustering, Pair Classification (P-CLS), Reranking, STS, Summarization (Summ.)). 
MTEB provides two evaluation metrics: Mean (Task) and Mean (Task Type). Mean (Task) is calculated as the average performance across all tasks within the benchmark. Mean (Task Type) is computed by first averaging the results within each task category and then averaging across all categories. 


\heading{MTEB (Code, v1)} 
This specialized benchmark focuses on evaluating embedding models in the context of software engineering. It covers code retrieval tasks across a wide array of popular programming languages, structured into 12 datasets. 

\heading{RTEB (Retrieval Text Embedding Benchmark, Beta)}
This benchmark focuses on retrieval tasks within high-stakes, specialized domains such as legal, financial, healthcare, multilingual, and code. 
It includes both open and closed datasets, offering a robust framework for evaluating real-world applicability. We use the open datasets, including 16 datasets, for our evaluation. 

Moreover, different tasks within each benchmark use distinct evaluation metrics. For example, retrieval tasks use NDCG@10 \cite{jarvelin2002cumulated}, classification tasks use Accuracy, and reranking tasks use MAP@1000.
Each dataset may contain multiple subsets, and the score for each dataset is obtained by averaging the scores across its subsets.
The final score for each benchmark is then calculated by averaging the scores across different tasks or datasets, which makes it less suitable for significance testing. 


\subsection{Implementation Details}
For the LLM backbone, we adopted Qwen3-0.4B and Qwen3-4B \cite{yang2025qwen3technicalreport} as the backbone for our framework. 
We fine-tune the Qwen3-0.6B and Qwen3-4B models \cite{yang2025qwen3technicalreport} using the contrastive loss for a single epoch on the Eng-Text-Data and General-Full-Data datasets, respectively. The training is on a machine with 8× Nvidia A800 (80GB) GPUs. 
For the larger Qwen3-4B model, we employ Low-Rank Adaptation (LoRA) \cite{edwardlora2021} for efficient fine-tuning, with the LoRA rank and alpha both set to 32. The learning rate is set to 5e-5, and the batch size is 128. For the smaller Qwen3-0.6B model, we apply full-parameter tuning with a learning rate of 5e-5 and a batch size of 256. 
The following settings are shared between both models: each dataset incorporates 7 hard negatives, which are provided by \citet{li2024making} and \citet{suresh2024cornstack}, the maximum sequence length is limited to 512 tokens, and the same dataset is used consistently within each training step. 
\vspace{-2mm}
\section{Comparisons of Different Training Strategies}
\label{sec:rq1}

Given the $K$ datasets $\{D_1, D_2, ..., D_K\}$ from multiple tasks, we summarize and investigate methods for training a single, general-purpose text embedding model. 
For the training function of the embedding model, we employ the standard InfoNCE loss function \cite{izacard2021unsupervised}: 
\begin{equation}
  L = -\log\frac{\exp(s(q,d^+))}{\exp(s(q,d^+)) + \sum \exp(s(q,D^-))},
\end{equation} 
where $D^{-}$ denotes the set of negative documents, and $s(q, d^+)$ is the scoring function—specifically, the cosine similarity between the query and document in our case. 
The use of in-batch negatives has been demonstrated to be highly effective for training dense embedding-based retrievers. However, applying the in-batch negatives approach to classification or clustering tasks can mislead the embedding model, as the ``passages'' within a mini-batch may belong to the same class and do not constitute true negatives. To address this, we adopt different negative sampling strategies for different tasks. For non-retrieval tasks (e.g., classification and clustering), $D^{-}$ contains only hard negative documents. For retrieval tasks, $D^{-}$ includes both hard negatives and in-batch negatives. For the training data, we use the \engdata dataset in this section. 


\begin{table}[t]
  \centering
  \small
  \renewcommand{\arraystretch}{0.9}
   \setlength\tabcolsep{1.8pt}
  \caption{Average performance (\%) of general text embedding models on MTEB(Eng, v2) with Qwen3-4B, using various model merging methods at the task level. Bold indicates the best performance among the different model merging methods.}
    \begin{tabular}{llllllllll}
    \toprule
  \multicolumn{1}{l}{Task-level} & \multicolumn{1}{l}{Se(N)} & \multicolumn{1}{l}{MS(S)} & \multicolumn{1}{l}{MS(N)} & \multicolumn{1}{l}{Kar} & \multicolumn{1}{l}{TA(N)} & \multicolumn{1}{l}{T(S)} & \multicolumn{1}{l}{TS(N)} & \multicolumn{1}{l}{SCE} & \multicolumn{1}{l}{Stock} \\
    \midrule
    Classification & 80.53  & \textbf{81.44}  & 80.52  & 81.43  & 80.55  & 80.53  & 79.70  & 81.06  & 61.23  \\
    Clustering & 46.47  & \textbf{52.53}  & 46.16  & 52.35  & 46.38  & 46.17  & 45.03  & 50.88  & 37.57  \\
    P-CLS & \textbf{81.95}  & 76.46  & 82.06  & 76.54  & \textbf{81.95}  & \textbf{81.95}  & 81.26  & 79.30  & 37.56  \\
    Reranking & 48.60  & 47.90  & \textbf{48.63}  & 47.84  & 48.58  & 48.58  & 48.26  & 47.99  & 35.14  \\
    Retrieval & 57.18  & 46.73  & \textbf{57.23}  & 46.75  & 57.17  & 57.17  & 57.24  & 53.14  & \phantom{0}4.80  \\
    STS   & 81.23  & 76.68  & \textbf{81.26}  & 76.74  & 81.22  & 81.22  & 80.86  & 74.43  & 37.39  \\
    Summ. & 32.35  & \textbf{38.46}  & 32.09  & 38.09  & 32.30  & 32.30  & 31.84  & 37.67  & \phantom{0}7.70  \\
    \midrule
    Mean(Task) & \textbf{65.71}  & 63.24  & 65.67  & 63.21  & 65.69  & 65.65  & 65.13  & 64.11  & 33.31  \\
    \bottomrule
    \end{tabular}%
  \label{tab:different_merging}%
\end{table}%

 \begin{table*}[t]
  \centering
   \renewcommand{\arraystretch}{0.9}
   \setlength\tabcolsep{3.0pt}
  \caption{Average performance (\%) of general text embedding models using different training strategies on MTEB (Eng, V2). \textbf{Bold} means the best performance on the task among the same LLM. ``(Number)'' indicates the number of datasets.}
    \begin{tabular}{lllllll|llllll}
    \toprule
    \multicolumn{1}{c}{\multirow{3}[5]{*}{AVERAGE}} & \multicolumn{6}{c}{Qwen3-0.6B}                & \multicolumn{6}{c}{Qwen3-4B} \\
\cmidrule(r){2-7}  \cmidrule(r){8-13}    \multicolumn{1}{c}{} & \multicolumn{4}{c}{\textbf{Data Scheduling}} & \multicolumn{2}{c|}{\textbf{Model Merging}} & \multicolumn{4}{c}{\textbf{Data Scheduling}} & \multicolumn{2}{c}{\textbf{Model Merging}} \\
\cmidrule(r){2-5}  \cmidrule(r){6-7}  \cmidrule(r){8-11} \cmidrule(r){12-13}   \multicolumn{1}{c}{} & \multicolumn{1}{l}{Batch} & \multicolumn{1}{l}{Dataset} & \multicolumn{1}{l}{Task} & \multicolumn{1}{l}{Two Stage} & \multicolumn{1}{l}{Dataset} & \multicolumn{1}{l|}{Task} & \multicolumn{1}{l}{Batch} & \multicolumn{1}{l}{Dataset} & \multicolumn{1}{l}{Task} & \multicolumn{1}{l}{Two stage} & \multicolumn{1}{l}{Dataset} & \multicolumn{1}{l}{Task} \\
\midrule
    Classification (8) & \textbf{86.01}  & 51.60  & 76.97  & 85.95  & 61.16  & 74.99  &  \textbf{87.28}  & 70.70  & 79.87  & 87.09  & 69.27  & 80.52  \\
    Clustering (8) & \textbf{54.68}  & 26.68  & 45.19  & 54.18  & 40.53  & 44.13  &  \textbf{59.05}  & 38.68  & 48.53  & 58.88  & 39.02  & 46.16  \\
    Pair Classification (3) & \textbf{81.20}  & 55.42  & 80.57  & 79.96  & 44.08  & 79.80  &  \textbf{82.69}  & 60.40  & 82.27  & 82.51  & 53.82  & 82.06  \\
    Reranking (2) & 45.81  & 35.86  & 45.16  & \textbf{45.87}  & 36.83  & 46.19  &  \textbf{49.46}  & 40.86  & 48.09  & 49.22  & 38.78  & 48.63 \\
    Retrieval (10) & 50.60  & 20.80  & 50.86  & 46.23  & 23.42  & \textbf{53.32}  &  55.03  & 31.55  & 57.10  & 51.51  & 32.72  & \textbf{57.23}  \\
    STS (9) & 78.34  & 49.26  & 77.29  & 77.46  & 45.24  & \textbf{78.47} &  81.10  & 60.62  & 81.28  & 81.11  & 52.60  & \textbf{81.26} \\
    Summarization (1) &  31.70 & 25.22  & \textbf{31.86}  & 30.85  & 24.58  & 31.25  &  35.29  & 28.30  & 32.73  & \textbf{37.18}  & 23.61  & 32.09  \\
    \midrule
    Mean (Task) &  \textbf{65.94}  & 37.58  & 62.08  & 64.46  & 41.11  & 62.33  &  \textbf{69.10}  & 49.45  & 65.99  & 68.19  & 47.06  & 65.67  \\
    Mean (TaskType) &  \textbf{61.19}  & 37.83  & 58.27  & 60.07  & 39.40  & 58.31  &  \textbf{64.27}  & 47.30  & 61.41  & 63.93  & 44.26  & 61.13 \\
    \bottomrule
    \end{tabular}%
  \label{tab:different_training}%
\end{table*}%
\subsection{Multi-Task Training Strategies}
We conduct a systematic study of multi-task training
strategies from two perspectives: data scheduling and model merging. 

\heading{Data Scheduling Strategies} We systematically compare the following representative training paradigms:
\begin{itemize}[leftmargin=*,itemsep=0pt,topsep=0pt,parsep=0pt]
\item \textbf{Batch-Level Shuffling (BLS)}: Following BGE-en-ICL \cite{li2024making} and GTE \cite{li2023towards}, each training batch is constructed by sampling data from a single dataset. These batches are then shuffled throughout training.  This approach aims to simulate a diverse distribution at a fine granularity during training. 
\item \textbf{Dataset-Level Sequential Training}: Models are trained on one complete dataset at a time, sequentially progressing through the tasks in the following order: classification, clustering, STS, and finally retrieval. Within each task, the order of datasets is random, except for retrieval. For retrieval, we trained on each dataset and found that MS MARCO document and MS MARCO Passage achieved the best and second-best performance, respectively. Therefore, the last two datasets used for retrieval are the MS MARCO document and the MS MARCO passage, while the others are arranged randomly. 
\item \textbf{Task-Level Sequential Training}: Within a single task type (e.g., retrieval), a batch-Level mixing strategy is applied to all datasets belonging to that task. Training proceeds sequentially across different task types (i.e., classification, clustering, sts, and then retrieval). 
\item \textbf{Two-Stage Training}: \citet{lee2024nv} proposed a two-stage training method. Specifically, the first stage involves contrastive training on retrieval-style datasets using in-batch negatives and curated hard-negative examples. The second stage performs contrastive learning on a mixture of sampled retrieval datasets and all non-retrieval datasets (e.g., classification and clustering tasks), this time without the use of in-batch negatives.
\end{itemize}

\heading{Model Merging Techniques}
Considering the model merging approach, we conduct training and merging from two perspectives: 
\begin{itemize}[leftmargin=*,itemsep=0pt,topsep=0pt,parsep=0pt]
  \item \textbf{Dataset-Level Merging}: For each dataset $D_i$, we individually train a model $M_i$, and then merge the $N$ models into a single model using the model merging strategy; and 
  \item \textbf{Task-Level Merging}: For each task, we train a model using the batch-level mixing training method, and integrate the N task-specific models into one unified model. 
\end{itemize}

We empirically evaluate seven model merging algorithms: SLERP (\textbf{Se}), Multi-SLERP (\textbf{MS}), Karcher mean (\textbf{Kar}), Task Arithmetic (\textbf{TA}), TIES-Merging (\textbf{TS}), SCE, and Model Stock (\textbf{Stock}). 
During the merging process, we employ two weighting schemes ($\{\alpha\}_{i=1}^N$): (1) weighting by the size of each model’s training dataset \textbf{(N)} and (2) equal weighting across all models \textbf{(S)}. Table \ref{tab:different_merging} presents the task-level results of the Qwen3-4B model on the MTEB(Eng, v2) dataset using different merging methods. 
For task-level merging using SLERP, which merges only two models at a time, we sequentially combine models trained on STS, clustering, classification, and retrieval tasks. 
The analysis indicates that: (1) The choice of model merging algorithm impacts the performance of the final merged model; (2) Multi-SLERP, Task Arithmetic, and TIES-Merging yield comparable performance, while Model Stock—a recently proposed merging method—produces the worst performance in the merged embedding model. This suggests that merging methods developed for different domains may not be well-suited for general text embedding models. To ensure consistency across all subsequent merging experiments, we consistently employ Multi-SLERP as the model merging method.

\begin{figure*}
    \centering
    \begin{subfigure}[b]{0.3\linewidth}
        \centering
        \includegraphics[width=\linewidth]{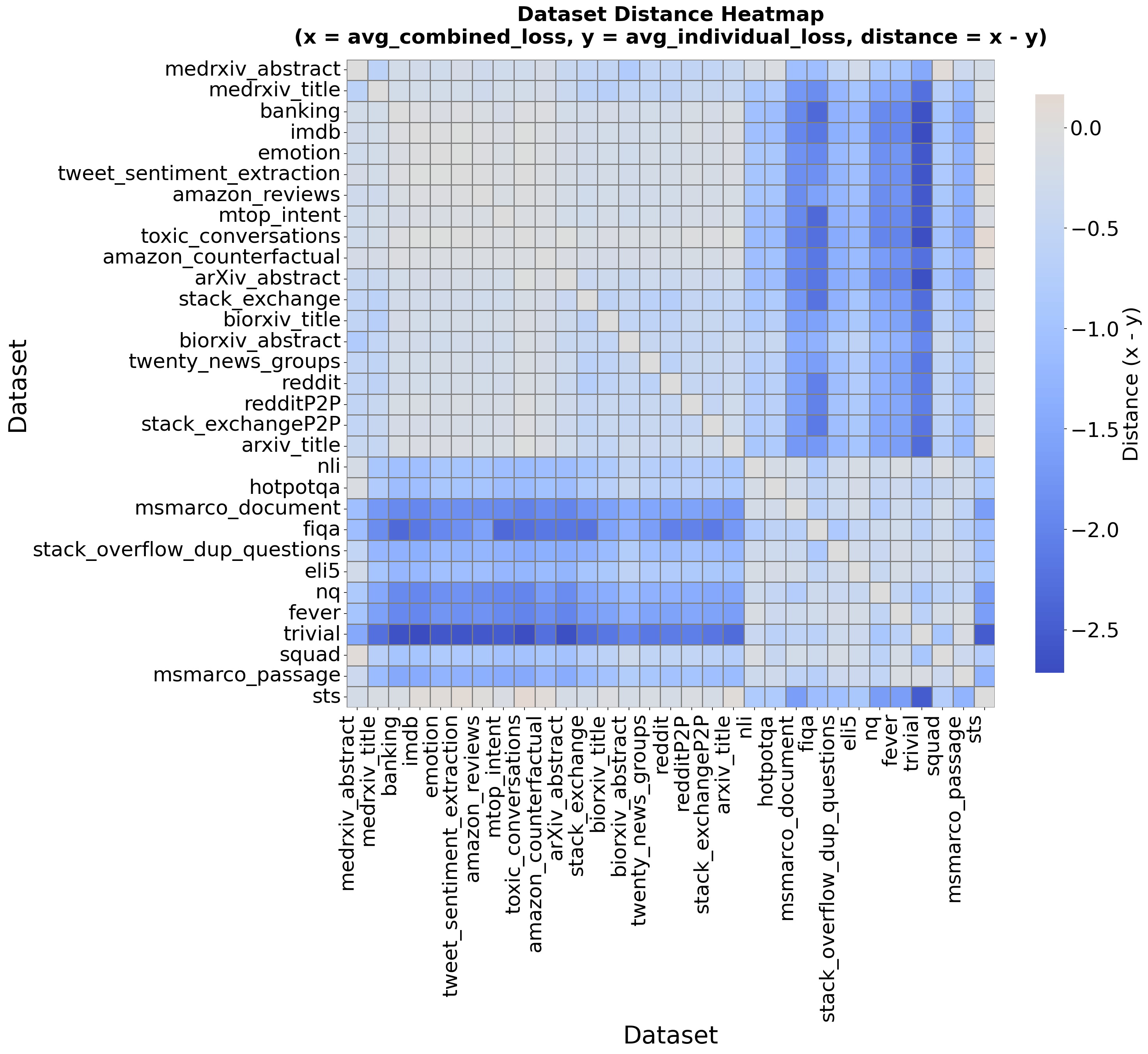}
        \caption{Pairwise Dataset Interaction Matrix.}
        \label{fig:heatmap}
    \end{subfigure}
    \hfill
    \begin{subfigure}[b]{0.68\linewidth}
        \centering
        \includegraphics[width=\linewidth]{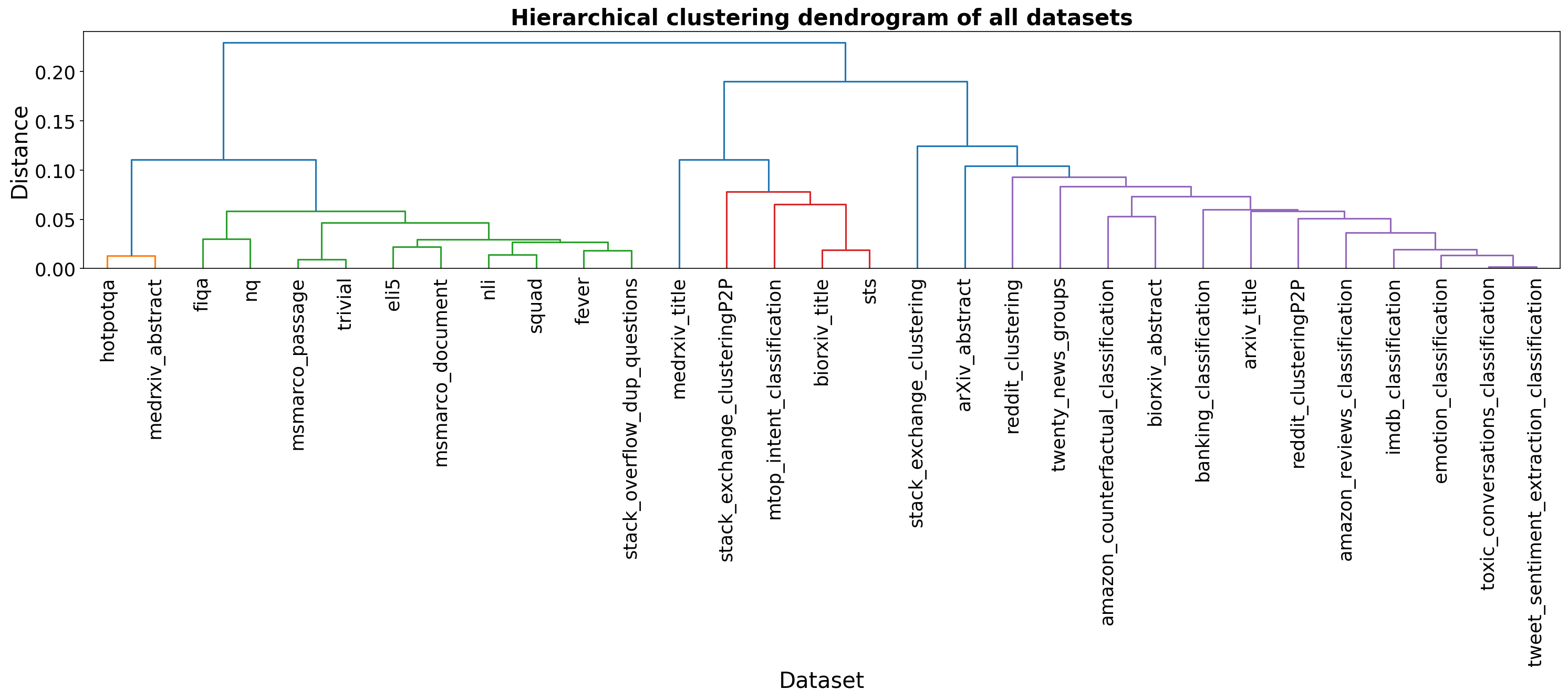}
        \caption{Hierarchical Clustering of Datasets.}
        \label{fig:dendrogram}
    \end{subfigure}
    \caption{(a) Difference between average joint and individual training losses for dataset pairs; (b) hierarchical clustering results.}
    \label{fig:combined_heatmap}
\end{figure*}

\vspace{-2mm}
\subsection{Experimental Results}
Table \ref{tab:different_training} shows the performance of different training strategies on MTEB (Eng, v2). 
We can observe the following findings:
(1) 
Across both model sizes (Qwen3-0.6B and Qwen3-4B), the batch-level shuffling strategy consistently achieves the highest average scores (65.94\% and 69.10\%, respectively). This strongly supports the hypothesis that fine-grained, interleaved exposure to diverse data within a batch is highly effective in mitigating catastrophic forgetting and fostering robust, generalizable representations. 
(2) 
The dataset-level strategy performs the worst across all metrics (37.58\% and 49.45\% averages), significantly lagging behind other methods on mean scores. This stark contrast validates the initial concern: sequential training on entire datasets leads to severe catastrophic forgetting, drastically impairing the model's ability to retain knowledge from previously seen data.
Moreover, the performance trend from dataset-level to task-level and finally to batch-level shuffling demonstrates a clear positive correlation between the granularity of data intermixing and final model capability. Task-level sequential shows a marked improvement over dataset-level, but is consistently outperformed by batch-level shuffling. This stepwise enhancement underscores the importance of frequent and fine-grained cross-dataset interactions during training for optimal knowledge integration. 
(3) 
The second stage's exclusive reliance on hard negatives likely forfeits the beneficial, diverse supervisory signals provided by in-batch negatives, which are crucial for learning high-quality retrieval representations. This highlights a potential task-dependent limitation of hard-negative-only fine-tuning phases.
(4) A novel observation from the results is that model merging strategies generally fail to outperform the data scheduling strategy (batch-level shuffling). 
This suggests that dynamically interleaving data during a single training phase (batch-level) is a more effective method for general text embedding models than statically averaging parameters from separately trained models after the fact. 

\vspace{-2mm}
\subsection{Cluster-Level Merging Method} 
To move beyond a heuristic combination of experts and towards a principled merging framework, we first seek to quantitatively determine whether pairs of training datasets exhibit a synergistic or conflicting relationship during multi-task training. 
The core hypothesis is that datasets promoting similar or complementary feature representations will yield lower training loss when combined, whereas those with conflicting learning signals will impair optimization. 
To systematically test this, we establish the following experimental protocol:
\begin{itemize}[leftmargin=*,itemsep=0pt,topsep=0pt,parsep=0pt]
    \item \textbf{Data Sampling for Controlled Comparison}: To isolate the effect of dataset composition from that of data volume, we randomly sample a fixed subset $D'_i$ with $N_{least}$ (the minimum dataset size among all training datasets) examples from each original training dataset $D_i$. 
    \item \textbf{Pairwise Joint Training Experiment}: We conduct an exhaustive pairwise training study. For every unique pair of datasets $(D'_i, D'_j)$, we train a model using a standard batch-level shuffling strategy on the combined set $D'_i \cup D'_j$. This results in $\frac{N \times (N-1)}{2}$ joint training runs. Additionally, we train $N$ individual models, each on a single subset $D'_i$, to establish baseline performance. 
    \item \textbf{Quantifying Pairwise Interaction}: For each pair $(i, j)$, we define a key metric: the average step loss over the course of training, denoted as $LC_{ij}$. We compare this to the average of the losses from the two corresponding individually trained models, $LA_{ij} = (\text{loss}_i + \text{loss}_j)/2$. 
    The pairwise difference $\delta_{ij} = \text{avg\_combined\_loss}_{ij} - \text{avg\_individual\_loss}_{ij}$ for all pairs forms a dataset interaction matrix, visualized in Figure \ref{fig:combined_heatmap} (a). 
    If $\delta_{ij} <0 $, the distance of the two datasets is $d_{ij}=1.0 - \frac{LC_{ij} }{LA_{ij}}$. Conversely, $d_{ij}=1.0 + \frac{\delta_{ij}}{LA_{ij}}$. To achieve flexible clustering, we employ hierarchical clustering \cite{murtagh2012algorithms} on all pairs, as shown in Figure \ref{fig:combined_heatmap} (b).  
\end{itemize}

\begin{figure}
    \centering
    \includegraphics[width=\linewidth]{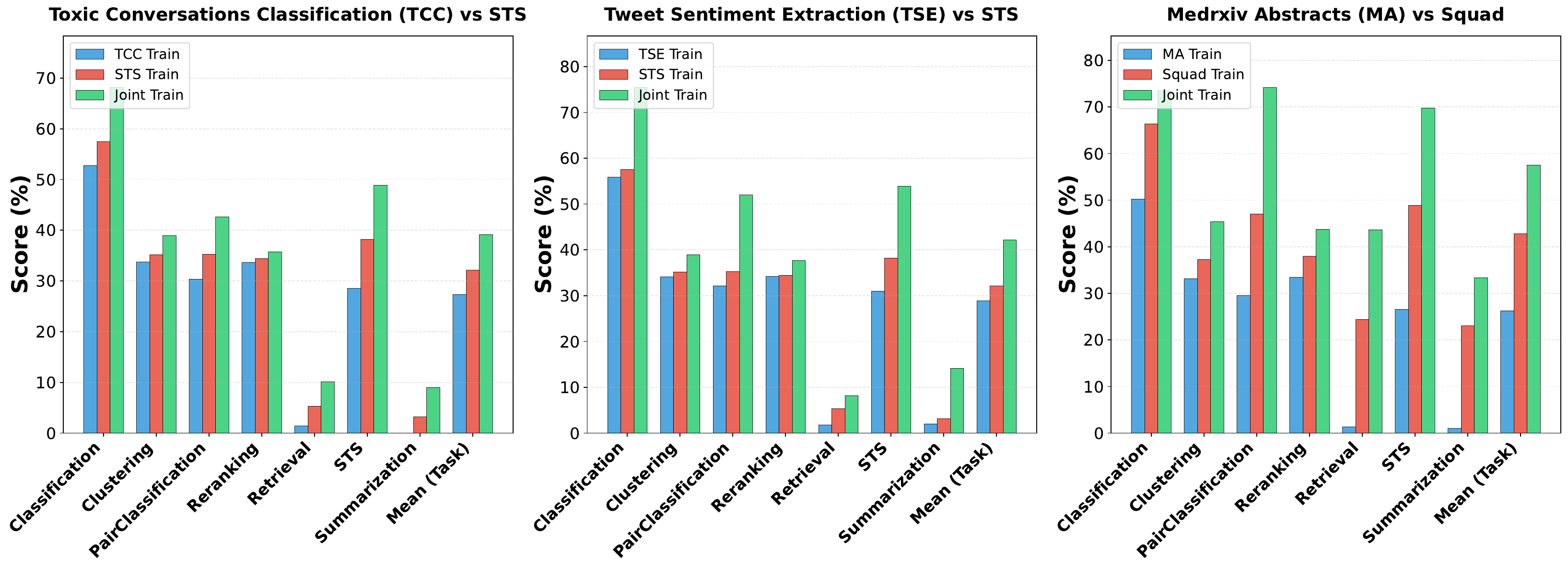}
    \caption{Average performance (\%) comparison on MTEB (Eng, v2) between models trained jointly and independently on three pairs of datasets.}
    \label{fig:independent_joint}
\end{figure}

From Figure \ref{fig:combined_heatmap} (a), we can find that only a small subset of dataset pairs (approximately 2\% of the total) are identified with a potential conflict $(\delta_{ij} > 0)$. 
We used the top-3 pairs with the highest $\delta_{ij}$ for in-depth analysis. 
For each selected pair, we trained independent training models and a joint training model using the batch-level shuffling approach on the original training data. 
The results, presented in Figure \ref{fig:independent_joint}, clearly demonstrate the superiority of joint training. 
For all three pairs, the model trained with batch-level shuffling on the combined data consistently outperformed the better of the two individually trained models across nearly all task categories. 
This indicates that even for datasets identified as potentially conflicting by our loss-difference metric, joint optimization yields a more robust and generalizable embedding model. 

\begin{table}[t]
  \centering
  \small
   \renewcommand{\arraystretch}{0.9}
   \setlength\tabcolsep{1.0pt}
  \caption{Average performance of models merged from clusters defined at different hierarchical thresholds using Qwen3-4B on MTEB(Eng, v2). ``T'' denotes the threshold used in the dataset clustering method.}
    \begin{tabular}{llllllll}
    \toprule
  MTEB(Eng, v2)  &  Task(S)&  Task (N) &  T0.15 (S) & T0.15 (N) &T0.2 (S) & T0.2 (N) & T1.0  \\
\midrule
    Mean (Task)   & 63.24  & 65.67  & 66.58  & 66.54  & 67.88  & 66.71 & 69.10 \\
    \bottomrule
    \end{tabular}%
  \label{tab:clutering_cut}%
\end{table}%

Building upon the clustering results in Figure \ref{fig:combined_heatmap} (b),  we investigate how the granularity of dataset clustering influences the performance of the subsequent model merging pipeline. 
We experiment with three thresholds: (1) Threshold = 1.0: All datasets belong to a single cluster. (2) Threshold = 0.15: 3 distinct clusters. (2) Threshold = 0.2: 2 distinct clusters. 
The comprehensive evaluation results on the MTEB (English, v2) benchmark are presented in Table \ref{tab:clutering_cut}. Two key observations emerge: 
(1) The models created by merging cluster-specific methods consistently outperform the models merged based on predefined task categories. 
(2) Counter-intuitively, the overall performance tends to improve as the number of clusters to be merged decreases. The single-cluster scenario, which is equivalent to direct batch-level shuffling on all data, achieves the highest scores.  This trend strongly suggests that synergistic relationships are dominant and widespread across our entire training collection.

\section{Bagging-Based Robust Model Merging}
Conventional model merging strategies are primarily designed to mitigate task conflict by training specialized models and subsequently merging them to enhance overall performance and robustness. However, our empirical analysis (Section \ref{sec:rq1}) reveals that within the general text embedding training, datasets predominantly exhibit synergistic relationships. This pivotal insight motivates a fundamental rethinking of the merging paradigm: instead of isolating potential conflicts, how can we design an efficient, robust, and sustainable framework that actively leverages this widespread synergy? Inspired by the Bagging (Bootstrap Aggregating) \cite{DBLP:journals/ml/Breiman96b} technique from ensemble learning, we propose Bagging-Based Robust Model Merging (\modelname). 
\begin{figure}
    \centering
    \includegraphics[width=\linewidth]{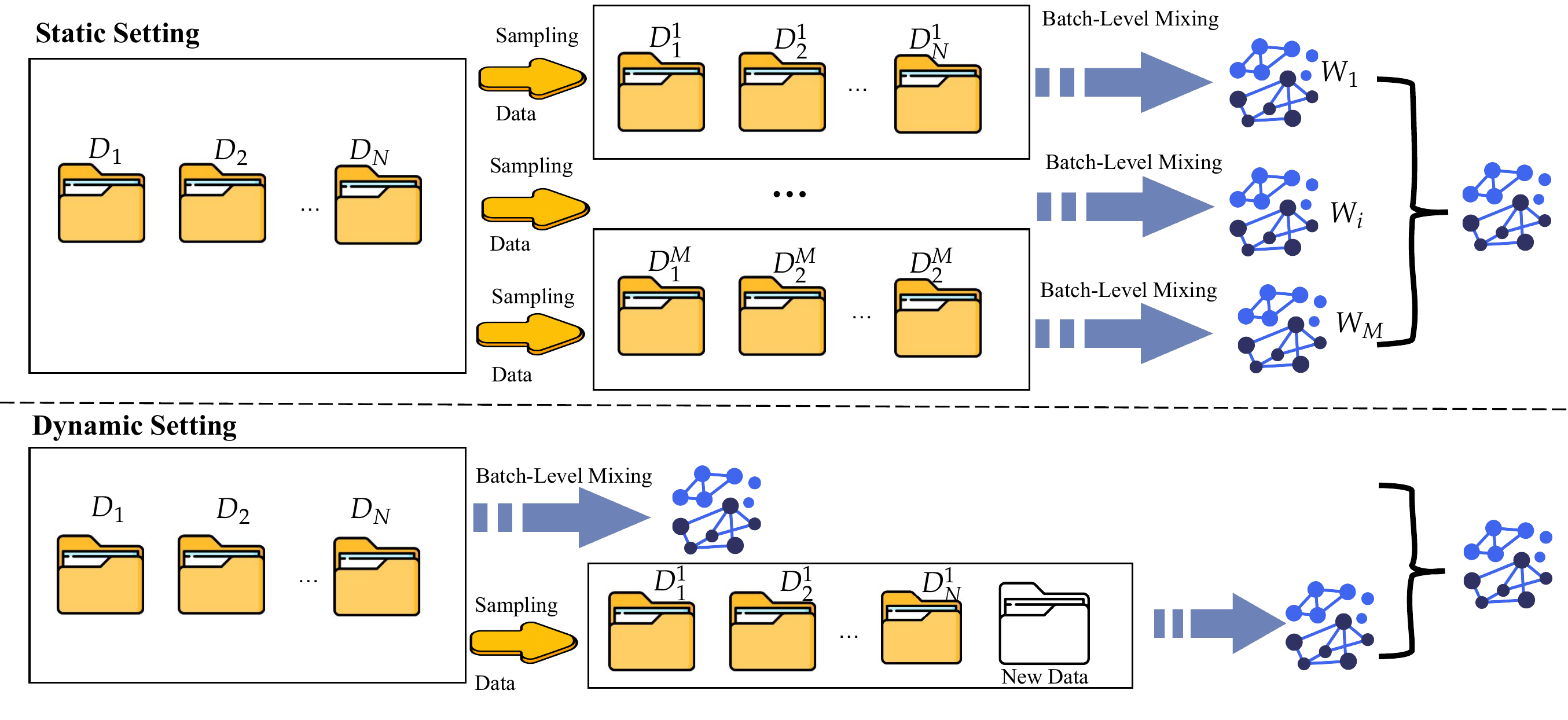}
    \caption{The overall framework of \modelname.}
    \label{fig:method}
\end{figure}

\subsection{Methodology}

\begin{table*}[t]
  \centering
  \small
     \renewcommand{\arraystretch}{0.9}
   \setlength\tabcolsep{1.8pt}
  \caption{Average performance comparison on different MTEB benchmarks under \engdata training datasets.   Bold and \underline{underline} indicate the best performance on each task across all models and within the same model scale, respectively. Since \engdata excludes the three training datasets, BGE-en-ICL’s IND and OOD are marked as ``-''. }
    \begin{tabular}{llccccccc|ccc|cc}
    \toprule
    \multicolumn{1}{c}{\multirow{2}[4]{*}{Size}} & \multicolumn{1}{c}{\multirow{2}[4]{*}{Method}} & \multicolumn{10}{c}{MTEB(Eng, v2)}                                           & \multicolumn{1}{l}{RTEB(beta)} & \multicolumn{1}{l}{MTEB(Code, v1)} \\
\cmidrule(r){3-12} \cmidrule(r){13-13}  \cmidrule(r){14-14}  \multicolumn{1}{c}{} &       & \multicolumn{1}{c}{Classification} & \multicolumn{1}{c}{Clustering} & \multicolumn{1}{c}{P-CLS} & \multicolumn{1}{c}{Reranking} & \multicolumn{1}{c}{Retrieval} & \multicolumn{1}{c}{STS} & \multicolumn{1}{c|}{Summ.} & \multicolumn{1}{c}{Mean (Task)} & \multicolumn{1}{c}{IND} & \multicolumn{1}{c|}{OOD} & \multicolumn{1}{c}{OOD} & \multicolumn{1}{c}{OOD} \\
    \midrule
    7B & BGE-en-ICL \cite{li2024making} & \textbf{88.78}  & 	57.80  & 	\textbf{85.39}  & 	48.02   & 	55.10 	&  \textbf{82.21}   & 	32.20   &	69.46   &  -  &  - & 56.51 & 	59.67   \\
    \midrule
    \multirow{6}[3]{*}{0.6B} & BLS & 86.01  & 54.68  & 81.20  & 45.81  & 50.60  & 78.34  & 31.70  & 65.94  & 73.09  & 56.80  & 46.61  & 56.12  \\
\cmidrule{2-14}          & \multicolumn{13}{l}{\modelname} \\
 &  $w/$ \{50\}-and-R & 85.34  & 55.07  & 80.97  & 46.84  & 53.07  & 79.19  & 31.11  & 66.69  & 73.27  & 58.29  & 52.52  & 61.10  \\
  &$w/$ \{60,60,60\} & 85.88  & 55.35  & \underline{82.08}  & \underline{47.37}  & 53.16  & 79.52  & 31.31  & 67.06  & 73.63  & 58.67  & 52.62  & 61.10  \\
 & $w/$ \{20, 40, 60,80, 100\} & \underline{86.13}  & \underline{55.60}  & 81.99  & 47.16  & \underline{53.31}  & \underline{80.16}  & 32.93  & \underline{67.36}  & \underline{73.98}  & \underline{58.89}  & \underline{53.20}  & \underline{61.12}  \\
 \midrule
    \multirow{6}[3]{*}{4B} & BLS & 87.28  & \textbf{\underline{59.05}}  & 82.69  & 49.46  & 55.03  & 81.10  & 35.29  & 69.10  & 75.78  & 60.57  & 60.52  & 65.19  \\
\cmidrule{2-14}          & \multicolumn{13}{l}{\modelname} \\
 & $w/$ \{50\}-and-R & 86.76  & 58.58  & 83.13  & 49.15  & 54.77  & 81.48  & 34.26  & 68.92  & 75.48  & 60.53  & 61.34  & 66.54  \\
  &  $w/$ \{60,60,60\} & \underline{87.38}  & 58.77  & 83.46  & 49.25  & 55.04  & \underline{81.90}  & 36.47  & 69.32  & 75.87  & 60.95  & 61.27  & 66.67  \\
 & $w/$ \{20, 40, 60,80,100\} & 87.34  & 59.00  & \underline{83.71}  & \underline{\textbf{49.52}}  & \textbf{\underline{55.67}}  & 81.95  & \underline{36.72}  & \textbf{\underline{69.56}}  & \textbf{\underline{75.97}} & \textbf{\underline{61.37}} & \textbf{\underline{61.56}}  & \textbf{\underline{66.74}}  \\
 \bottomrule
    \end{tabular}%
  \label{tab:static_results}%
\end{table*}%

Traditional bagging improves robustness by training multiple models on bootstrapped samples and aggregating their predictions, but deploying multiple models incurs high computational overhead for embedding-based services. By merging parameters from multiple models into a single embedding model, we retain ensemble benefits without extra inference cost. 
Inspired by bagging, we design \modelname~for two scenarios: a static setting (fixed corpus) and an incremental learning setting (continuous data integration). In the incremental setting, when new data arrives, we train a lightweight update model on the new data plus a small sample of historical data, then merge it with the existing model—efficiently incorporating new knowledge while mitigating forgetting.
The overall framework of \modelname is illustrated in Figure \ref{fig:method} and detailed below.

\heading{Static Setting}
Given a fixed, comprehensive training corpus $ \mathcal{D} = \{D_1, D_2, ..., D_N\} $, the goal is to produce a single robust embedding model. The process, depicted in the upper part of Figure \ref{fig:method}, consists of two stages:
(1) Parallel Training on Diversified Sampled Data: We sampled \( M \) distinct subsets $\{\mathcal{D}^{(1)}, \mathcal{D}^{(2)}, ..., \mathcal{D}^{(M)}\}$ by sampling from $\mathcal{D}$ using $\mathcal{K}=
\{k_1, k_2,...,k_m\}$ ratios, respectively. 
Each dataset $D^m_i$ of subset $\mathcal{D}^{(m)}$ is sampled from each dataset $D_i$ independently, and then $\mathcal{D}^{(m)}$ is used to train an embedding model $M_m$ with parameter $ W_m $ independently using a standard batch-level shuffling. 
This yields a set of embedding models $\Theta = \{W_1, W_2, ..., W_M\}$. 
(2) Parameter-Space Fusion: The ensemble of models  is fused into a single, robust model $\theta_{\text{merged}}$ via parameter-space merging techniques:
    \begin{equation}
        \theta_{\text{static}} = \mathcal{M}(W_1, W_2, ..., W_M),
    \end{equation}
    where $\mathcal{M}(\cdot)$ denotes the merging function (e.g., Multi-SLERP). 
Moreover, to ensure the same training cost as BLM trained on the full corpus, we introduce a variant: \modelname with ``\{50\}-and-R’', which refers to models trained separately on a 50\% random sample and the remaining 50\% of the training data, and then merged.

\heading{Incremental Learning}
In the incremental learning scenario, we start with a pre-trained embedding $M_0$ with parameters $W_0$, trained on an original corpus $\mathcal{D}$. When a new collection of datasets $\mathcal{D}'$ arrives, the objective is to efficiently integrate this new knowledge without full retraining. The process, shown in the lower part of Figure \ref{fig:method}, is as follows:
(1) Construct a Representative Core Subset: From the original corpus $\mathcal{D}$, we randomly sample a representative subset $ \mathcal{D}_{\text{core}} \subset \mathcal{D}$ using a small ratio $k$. This subset preserves a snapshot of previously learned knowledge. A new embedding model $W_{\text{new}}$ is trained on the combination of the new data and the core subset using batch-level shuffling: 
    \begin{equation}
        W_{\text{new}} \leftarrow \text{Train}(\mathcal{D}' \cup \mathcal{D}_{\text{core}}).
    \end{equation}
(2) Merge the Old and New Model: The final model is obtained by merging the original model $W_0$ and the new model $W_{\text{new}}$:
    \begin{equation}
         W_{\text{dynamic}} = \mathcal{M}(W_0, W_{\text{new}}).
    \end{equation}
This dynamic approach dramatically reduces training cost compared to retraining on $\mathcal{D} \cup \mathcal{D}'$, as it requires training only on the (typically much smaller) set $\mathcal{D}' \cup \mathcal{D}_{\text{core}}$, followed by a low-cost merging operation. The core subset $\mathcal{D}_{\text{core}}$ acts as a regularizer, preventing catastrophic forgetting of essential prior knowledge during the training of $\theta_{\text{new}}$.

In summary, \modelname provides a unified, efficient framework that harnesses dataset synergy. 
It improves robustness in static settings via ensemble-based merging and enables sustainable, low-cost model updating in incremental learning settings, making it highly suitable for real-world embedding model development and maintenance.

\vspace{-2mm}
\subsection{Evaluation Setting} 
To rigorously evaluate the adaptability and generalization capabilities of our embedding model, we design two distinct experimental settings: \textbf{static} and \textbf{incremental learning}. 
The static setting utilizes \engdata as a fixed training data for general text embedding training. 
In contrast, the incremental learning setting incrementally augments the training data with additional resources, e.g., multilingual retrieval and code retrieval tasks data, building upon the initial \engdata foundation. 
This dynamic approach is intended to validate the efficiency and effectiveness of our incremental learning merging strategy, specifically in its ability to adapt the model to new domains without incurring the computational costs associated with full retraining. 
For our \modelname, the $k$ in the incremental learning setting is 40\%. 
To ensure comprehensive assessment, the embedding model is evaluated across three representative benchmarks: MTEB (English, v2), RTEB (beta), and MTEB (Code, v1), enabling us to examine performance across varied domains. 
Furthermore, our evaluation protocol encompasses two distinct test settings to systematically characterize model generalization: 
\textbf{In-Domain (IND)}: Test data is drawn from the same distribution as the training data, measuring standard performance.
\textbf{Out-of-Distribution (OOD)}: Test data are entirely absent from the training data, thus probing the model’s robustness to novel retrieval tasks and its capacity for cross-task generalization. 

\vspace{-2mm}
\subsection{Experimental Results}

\heading{Static Setting}
The results are summarized in Table \ref{tab:static_results}. We can find that:
(1) Our \modelname variants consistently outperform the standard batch-level shuffling baseline across both Qwen3-0.4B and Qwen3-4B models on both in-domain performance and OOD generalization performance. 
Notably, the ``\{50\}-and-R'' variant—which incurs equivalent total training cost to batch-level shuffling—delivers better or comparable MTEB(eng, v2) performance alongside improved OOD scores (e.g., $\uparrow$5.9 on RTEB(beta) for the 0.6B model). demonstrating competitive and superior generalization of our \modelname paradigm. 
(2) Performance improves with increased diversity in the sampling ensemble. The ``\{20, 40, 60, 80, 100\}'' variant, which incorporates models trained on five different data partitions, achieves the best overall scores, setting a best performance of \textbf{69.56} on MTEB(Eng, v2) for the 4B model. 
(3) While BGE-en-ICL was trained on similar training data, we excluded three datasets from its training set because they lacked available training data on MTEB data. 
BGE-en-ICL employs in-context learning with few-shot and zero-shot variants. 
To ensure a fair comparison, we only evaluate its zero-shot variant. 
Although BGE-en-ICL achieves stronger performance on MTEB(Eng v1) compared to batch-level shuffling, it shows weaker out-of-distribution (OOD) generalization on RTEB and MTEB(Code, v1). 
Our \modelname variants—\{20, 40, 60, 80, 100\}—consistently outperform BGE-en-ICL across all three benchmarks, further indicating the superior generalization ability of our method. 

\begin{table}[t]
  \centering
   \renewcommand{\arraystretch}{0.8}
  \small
  \caption{Average performance (\%) of different models on the incremental learning setting. ``Eng'' and ``Gen'' means \engdata training data and \alldata training data, respectively.  \textbf{Bold} means the best performance among the same LLM. 
  }
    \begin{tabular}{llll|lll}
    \toprule
    \multicolumn{1}{c}{\multirow{3}[6]{*}{Benchmark}} & \multicolumn{3}{c}{Qwen3-0.6B} & \multicolumn{3}{c}{Qwen3-4B} \\
\cmidrule(r){2-4}  \cmidrule(r){5-7}          & \multicolumn{2}{c}{BLS} & \multicolumn{1}{l|}{\multirow{2}[4]{*}{\modelname}} & \multicolumn{2}{c}{BLS} & \multicolumn{1}{l}{\multirow{2}[4]{*}{\modelname}} \\
\cmidrule{2-3}\cmidrule{5-6}          & \multicolumn{1}{l}{Eng} & \multicolumn{1}{l}{Gen} &       & \multicolumn{1}{l}{Eng} & \multicolumn{1}{l}{Gen} &  \\
    \midrule
    \multicolumn{7}{c}{MTEB (Eng, v2)} \\
    \midrule
    Classification & 86.01 & 86.15 & \textbf{86.15} & 87.28 & \textbf{87.65} & 87.28 \\
    Clustering & 54.68 & 55.05 & \textbf{55.33} & 59.05 & \textbf{59.23} & 59.02 \\
    PairClassification & 81.2  & 78.59 & \textbf{82.06} & 82.69 & 82.31 & \textbf{83.85} \\
    Reranking & 45.81 & 45.55 & \textbf{46.95} & 49.46 & 49.09 & \textbf{49.73} \\
    Retrieval & 50.60  & 51.24 & \textbf{53.21} & 55.03 & \textbf{55.99} & 55.72 \\
    STS   & 78.34 & 76.37 & \textbf{79.69} & 81.10  & 80.90  & \textbf{82.17} \\
    Summarization & 31.7  & \textbf{34.29} & 33.99 & 35.29 & 35.70  & \textbf{36.44} \\
    \midrule
    Mean (Task) & 65.94 & 65.62 & \textbf{67.20}  & 69.10  & 69.36 & \textbf{69.63} \\
    Mean (Task Type) & 61.19 & 61.03 & \textbf{62.48} & 64.27 & 64.41 & \textbf{64.89} \\
    \midrule
    RTEB(beta) & 46.61 & 52.54 & \textbf{54.93} & 60.52 & 63.03 & \textbf{63.15} \\
    \midrule
    MTEB (Code, v1) & 56.12 & 62.29 & \textbf{64.46} & 65.19 & \textbf{70.0}    & 69.45 \\
    \bottomrule
    \end{tabular}%
  \label{tab:dynamic_results}%
\end{table}%

\heading{Incremental Learning Setting}
Our experiments in the incremental learning setting, as summarized in Table~\ref{tab:dynamic_results}, comprehensively demonstrate the effectiveness and efficiency of the proposed Bagging-Based Robust Model Merging (\modelname) framework for continual text embedding model development.
From the Table \ref{tab:dynamic_results}, we can find that:
(1) \modelname consistently delivers great and balanced improvements across all three evaluated benchmarks and both model scales. 
Notably, \modelname requires only 40\% of the original \engdata, combined with the incremental data from \alldata, rather than full retraining on the entire expanded dataset. Despite this reduced training data and cost, \modelname achieves substantial and consistent gains, most prominently for Qwen3-0.6B, where it not only recovers but surpasses the batch-level shuffling results on the three benchmarks. For Qwen3-4B, \modelname also matches or exceeds the strongest baselines across all benchmarks, demonstrating its robustness and scalability. 
(2) We observe that expanding the training data from the \engdata to the \alldata, (which includes English, multilingual, and code retrieval data),  leads to consistent performance improvements on the RTEB (beta) and MTEB (Code, v1) benchmarks for both Qwen3-0.6B and Qwen3-4B backbones. 
On the MTEB(Eng, v2) benchmark, the impact is different for LLM backbones: for Qwen3-0.6B, training on \alldata actually leads to a slight decrease in performance, while for Qwen3-4B, there is a marginal improvement. 
In summary, \modelname provides an effective and resource-efficient solution for continual model updating in real-world settings, successfully integrating new knowledge while preserving and enhancing overall embedding quality. This makes it highly practical for the sustainable development and maintenance of large-scale text embedding models in incremental learning environments.

\vspace{-2mm}
\subsection{Model Merging Variants}

\begin{table}[t]
  \centering
  \renewcommand{\arraystretch}{0.9}
   \setlength\tabcolsep{2.5pt}
  \caption{Average performance of different \modelname variants using Qwen3-0.6B trained on \engdata, evaluated on MTEB (English, v2). ``TC'' means the training cost using GPU hours compared to the batch-level shuffling method. }
    \begin{tabular}{lcccc}
    \toprule
    Dataset Name & TC & \multicolumn{1}{l}{Mean(Task)} & \multicolumn{1}{l}{IND} & \multicolumn{1}{l}{OOD} \\
    \midrule
    BLS & 1.0 & 65.94 & 73.09 & 57.28 \\
    \midrule
     \modelname $w/$ \{50\}-and-R & 1.0& 66.69 & 73.27 & 58.59 \\
     \modelname $w/$ \{60, 60\}& 1.2 & 66.37 & 73.15 & 57.93 \\
     \modelname $w/$ \{60, 80\}& 1.4 & 66.89 & 73.75 & 58.36 \\
     \modelname $w/$ \{50, 50, 50\}& 1.5 & 66.35 & 73.30  & 57.59 \\
     \modelname $w/$ \{60, 60, 60\}&1.8 & 67.06 & 73.63 &	58.67 \\
     \modelname $w/$ \{40, 60, 80\}& 1.8 & 67.00    & 73.64 & 58.76 \\
     \modelname $w/$ \{80, 100\}&1.8 & 67.20 &73.95  & 58.89  \\
     \modelname $w/$ \{20, 40, 60, 80\}& 2.0 & 66.99 & 73.55 & 58.83 \\
     \modelname $w/$ \{60, 80, 100\}& 2.4 & 67.06 & \textbf{74.06} & 58.90 \\
     \modelname  $w/$ \{40, 60, 80, 100\} &2.8 & 67.35 & 73.95 & \textbf{59.18} \\
     \modelname $w/$ \{20, 40, 60, 80, 100\}&3.0 & \textbf{67.36} & 73.98 & 59.14 \\
    \bottomrule
    \end{tabular}%
  \label{tab:different_bagging}%
\end{table}%

To examine the effect of varying the number and composition of merged models, we evaluate multiple bagging combinations for \modelname, as summarized in Table~\ref{tab:different_bagging}. 
Our results reveal several key findings:
(1) All bagging and merging strategies outperform simple batch-level shuffling on both in-domain and out-of-domain evaluations. 
This demonstrates the effectiveness and robustness of the \modelname approach in leveraging diverse training subsets.
(2) Increasing the training cost, which corresponds to merging more models, generally leads to higher performance across both in-domain and OOD benchmarks. Merging strategies that incorporate a model trained on the full dataset (e.g., 100\%) are crucial for maximizing performance. For instance, the configuration \{80, 100\} (TC=1.8) outperforms \{20, 40, 60, 80\} (TC=2.0) on both in-domain and OOD metrics, despite the latter having a higher training cost but lacking a fully-trained model in the merge.

\section{Conclusion and Future Work} 
In this work, we present a comprehensive analysis of joint training and model merging strategies for general-purpose text embedding models. 
Our systematic experiments reveal that task conflicts are infrequent and that batch-level mixing generally leads to mutual improvement across diverse NLP tasks. 
However, we identify a critical limitation of conventional joint training: diminished out-of-domain generalization. 
To address these challenges, we introduce \modelname, a robust model merging framework inspired by bagging, which samples data subsets for training multiple models and merges them into a single, scalable embedding model. 
This approach not only enhances generalization and reduces inference costs compared to traditional ensemble methods but also supports efficient incremental learning by enabling rapid integration of new data with minimal retraining. 
Extensive empirical evaluations on MTEB and related benchmarks demonstrate that \modelname achieves state-of-the-art performance in both in-domain and out-of-domain scenarios, and substantially improves the efficiency of continual adaptation. 
Our findings offer practical guidance for building robust, generalizable, and efficient text embedding models. 

For future work, critical directions include: (1) Our current approach leverages existing model merging methods such as Multi-SLERP, and our experiments reveal that the more recent model merging method may not be directly suitable for general text embedding tasks. 
Therefore, developing new merging methods tailored for general-purpose text embedding remains an important direction for future research.  
(2) In addition, assessing the effectiveness of our results on downstream applications—including retrieval-augmented generation. Such an evaluation will ensure that robust and generalizable text embeddings can be reliably deployed in evolving NLP environments.

\clearpage
\bibliographystyle{ACM-Reference-Format}
\balance
\bibliography{reference}

\end{document}